\documentclass[useAMS,usenatbib]{mn2e}
\usepackage{array}
\usepackage{times}
\usepackage{graphicx}
\usepackage{mathptmx}
\usepackage[usenames]{color}

\usepackage{hyperref}
\hypersetup{colorlinks=true,citecolor=blue,linkcolor=purple,filecolor=black,runcolor=black}
\begin{document}
\newcommand{\comment}[1]{}
\definecolor{purple}{RGB}{160,32,240}
\newcommand{\peter}[1]{\textcolor{purple}{(Peter: \bf #1)}}
\newcommand{\AK}[1]{\textcolor{red}{(AK: \bf #1)}}
\newcommand{\macc}{M_\mathrm{acc}}
\newcommand{\mpeak}{M_\mathrm{peak}}
\newcommand{\mnow}{M_\mathrm{now}}
\newcommand{\vacc}{v_\mathrm{acc}} 
\newcommand{\vpeak}{v_\mathrm{peak}} 
\newcommand{\vnow}{v^\mathrm{now}_\mathrm{max}}

\newcommand{\ahf}{\textsc{AHF}}
\newcommand{\rockstar}{\textsc{Rockstar}}
\newcommand{\hbt}{\textsc{HBT}}
\newcommand{\subfind}{\textsc{SubFind}}
\newcommand{\velociraptor}{\textsc{VELOCIraptor}}
\newcommand{\stf}{\textsc{STF}}

\newcommand{\Mnfw}{M_\mathrm{NFW}}
\newcommand{\Msun}{\mathrm{M}_{\odot}}
\newcommand{\mvir}{M_\mathrm{vir}}
\newcommand{\rvir}{R_\mathrm{vir}}
\newcommand{\vmax}{v_\mathrm{max}}
\newcommand{\vmac}{v_\mathrm{max}^\mathrm{acc}}
\newcommand{\mvac}{M_\mathrm{vir}^\mathrm{acc}}
\newcommand{\sfr}{\mathrm{SFR}}
\newcommand{\plotgrace}[1]{\includegraphics[angle=-90,width=\columnwidth,type=eps,ext=.eps,read=.eps]{#1}}
\newcommand{\plotgraceflip}[1]{\includegraphics[angle=-90,width=\columnwidth,type=eps,ext=.eps,read=.eps]{#1}}
\newcommand{\plotlargegrace}[1]{\includegraphics[angle=-90,width=2\columnwidth,type=eps,ext=.eps,read=.eps]{#1}}
\newcommand{\plotlargegraceflip}[1]{\includegraphics[angle=-90,width=2\columnwidth,type=eps,ext=.eps,read=.eps]{#1}}
\newcommand{\plotminigrace}[1]{\includegraphics[angle=-90,width=0.5\columnwidth,type=eps,ext=.eps,read=.eps]{#1}}
\newcommand{\plotmicrograce}[1]{\includegraphics[angle=-90,width=0.25\columnwidth,type=eps,ext=.eps,read=.eps]{#1}}
\newcommand{\plotsmallgrace}[1]{\includegraphics[angle=-90,width=0.66\columnwidth,type=eps,ext=.eps,read=.eps]{#1}}

\newcommand{\hinv}{h^{-1}}
\newcommand{\mpc}{\rm{Mpc}}
\newcommand{\hmpc}{$\hinv\mpc$}

\title[Halo Finding for Major Mergers]{Major Mergers Going Notts: Challenges for Modern Halo Finders}

\author[P.\ Behroozi et al.]
{Peter Behroozi$^{1}$,\thanks{email: behroozi@stsci.edu}
 Alexander Knebe$^{2,3}$,
 Frazer R.\ Pearce$^{4}$,
 Pascal Elahi$^{5}$,
 Jiaxin Han$^{6}$, \newauthor
 Hanni Lux$^{4,7}$,
 Yao-Yuan Mao$^{8,9}$,
 Stuart I.\ Muldrew$^{10}$,
 Doug Potter$^{11}$,
 Chaichalit Srisawat$^{12}$
 \\
 \\
$^{1}$Space Telescope Science Institute, Baltimore, MD 21218, USA\\
$^{2}$Departamento de F\'isica Te\'{o}rica, M\'{o}dulo 15, Facultad de Ciencias, Universidad Aut\'{o}noma de Madrid (UAM), 28049 Madrid, Spain\\
$^{3}$Astro-UAM, UAM, Unidad Asociada CSIC\\
$^{4}$School of Physics \& Astronomy, University of Nottingham, Nottingham NG7 2RD, UK\\
$^{5}$Sydney Institute for Astronomy, School of Physics, A28, The University of Sydney, NSW 2006, Australia\\
$^{6}$Institute for Computational Cosmology, Department of Physics, Durham University, South Road, Durham DH1 3LE, UK\\
$^{7}$Department of Physics, University of Oxford, Denys Wilkinson Building, Keble Road, Oxford, OX1 3RH, UK\\
$^{8}$Kavli Institute for Particle Astrophysics and Cosmology \& Physics Department, Stanford University, Stanford, CA 94305, USA\\
$^{9}$SLAC National Accelerator Laboratory, Menlo Park, CA, 94025, USA\\
$^{10}$Department of Physics and Astronomy, University of Leicester, University Road, Leicester, LE1 7RH, UK\\
$^{11}$Institute for Computational Science, University of Zurich, 8057 Zurich, Switzerland\\
$^{12}$Department of Physics and Astronomy, University of Sussex, Brighton BN1 9QH, UK\\
}

\setlength{\topmargin}{-1.2cm}

\date{Released \today}

\pagerange{\pageref{firstpage}--\pageref{lastpage}} \pubyear{2015}

\maketitle

\label{firstpage}

\begin{abstract}
Merging haloes with similar masses (i.e., major mergers) pose significant challenges for halo finders.  We compare five halo finding algorithms' (\ahf, \hbt, \rockstar, \subfind, and \velociraptor) recovery of halo properties for both isolated and cosmological major mergers.  We find that halo positions and velocities are often robust, but mass biases exist for every technique.  The algorithms also show strong disagreement in the prevalence and duration of major mergers, especially at high redshifts ($z>1$).  This raises significant uncertainties for theoretical models that require major mergers for, e.g., galaxy morphology changes, size changes, or black hole growth, as well as for finding Bullet Cluster analogues.  All finders not using temporal information also show host halo and subhalo relationship swaps over successive timesteps, requiring careful merger tree construction to avoid problematic mass accretion histories.  We suggest that future algorithms should combine phase-space and temporal information to avoid the issues presented.
\end{abstract}
\begin{keywords}
methods:numerical -- dark matter -- galaxies: haloes
  \end{keywords}

\section{Introduction}

\label{s:introduction}

In the Lambda Cold Dark Matter paradigm, nonlinear gravitational collapse of matter overdensities yields self-bound structures known as ``haloes.''  Smaller haloes merge onto larger ones continuously, and are called ``subhaloes'' as long as they remain distinguishable within the radius of the larger halo.  Major mergers -- i.e., mergers between two haloes of similar mass -- occur rarely.  Estimates from recent simulations suggest that haloes at $z=0$ experience (on average) one merger per halo per 10 Gyrs \citep{Fakhouri08,Wetzel09,BWC13}.

Despite their infrequency, major mergers have been invoked to explain a surprisingly broad range of galaxy phenomena.  Galaxy growth correlates tightly with halo growth \citep[see][for recent constraints]{Leauthaud12,Wang12,Behroozi13,BWC13,Moster12,BehrooziHighZ}, so the significant disturbance to haloes in major mergers could also imply significant changes in observable galaxy properties.  Merger-linked phenomena with significant recent interest include active galactic nuclei (AGN) activity \citep{Kocevski12,Newton13} and associated black hole growth \citep{Treister12,Bonoli12}, Ultra-Luminous InfraRed Galaxy (ULIRG) triggering \citep{Kartaltepe10,Draper12}, galaxy morphology and size changes \citep{Bernardi11,Prieto13}, galaxy number density changes \citep{Lotz11,BehrooziND}, velocity dispersion evolution \citep{Oser12}, star formation quenching/triggering \citep{Kaviraj13,BehrooziQ}, galactic winds \citep{Hopkins13,Rupke13}, buildup of intracluster light \citep{Laporte13}, buildup of spheroidal bulges \citep{Sales12,Wilman13}, dispersal of magnetic fields \citep{Xu10}, and creation of tidal shells \citep{Wang-Tidal12}. 
 They also represent an important systematic for cluster analysis, including violations of hydrostatic equilibrium for X-ray masses \citep{Akahori09,Takizawa10,Bourdin11,Nelson12}, biases in Sunyaev-Zel'dovich signals \citep{Rudd09,SZ11,Krause12}, and incidence of Bullet Cluster-like systems \citep{Thompson14,Bouillot15}.

Predicting how major mergers will impact observables often involves a dark matter or hydrodynamical simulation \citep[see][for a review]{Kuhlen12}, a halo finder to convert the simulation particle data into a list of haloes and their properties \citep[see][for reviews]{Knebe11,Knebe13}, a merger tree algorithm to connect haloes across redshifts \citep[see][for a review]{Srisawat13}, and optionally a theoretical model for galaxy formation \citep[see][for a review]{Somerville14}.  The role of the halo finder in this process has been investigated in a recent series of workshops (including ``Haloes Going MAD,'' ``Subhaloes Going Notts,'' and ``Sussing Merger Trees'') and papers \citep{Knebe11,Onions12,Elahi13,Onions13,Pujol13,Knebe13b,Knebe13,Srisawat13,Perez13,Lee14,Hoffmann14}.  This paper, arising out of the ``Subhaloes Going Notts'' and ``Sussing Merger Trees'' workshops, continues this pattern with an investigation into how halo finders treat major mergers.

Halo finder recovery of very minor subhaloes (mass ratios $<$1:10) has already been investigated \citep{Muldrew11,Knebe11,Knebe13,Onions12}.  Finders which use particle positions alone to initially classify subhaloes are able to perform just as well as finders which use additional information (e.g., particle velocities or historical positions) in the outer halo, as long as gravitational unbinding is performed \citep{Onions12,Knebe13}.  As the larger ``host'' halo has a much larger velocity dispersion than the smaller subhalo, particle binding energies can very effectively distinguish between particles belonging to the host halo and to the subhalo.  In major mergers (mass ratios greater than 1:3), the host and the subhalo have similar velocity dispersions, making particle assignment much harder.  Additionally, the choice of which halo to call the ``host halo'' and which to call the ``subhalo'' can be ambiguous for major mergers, and can change over time unless temporal information is used \citep{Tweed09,HBT,Rockstar,Srisawat13}.  Hence, we investigate halo finders' abilities not only to recover halo properties in major mergers but also to follow halo properties smoothly across simulation timesteps.

We divide the results into several sections.  In \S \ref{s:desc}, we briefly describe the participating halo finders.  We describe ``static'' tests of halo finding, with overlapping mock \cite{NFW97} (NFW) profiles, in \S \ref{s:static}.  ``Dynamical'' tests are presented in \S \ref{s:dynamic}, where two mock NFW profiles are allowed to merge in an isolated simulation.  We present tests drawn from cosmological simulations in \S \ref{s:cosmological}.  Finally, we discuss the impact of these results in \S \ref{s:discussion} and summarize our conclusions in \S \ref{s:conclusions}.  Throughout this work, halo masses are calculated as spherical overdensities, and host halo masses include all substructure masses.

\section{Common Terms and Halo Finder Descriptions}
\label{s:desc}

In this section, we define common terms and briefly describe the participating halo finders (\ahf, \hbt, \rockstar, \subfind, and \velociraptor).  These descriptions include the overall algorithm employed and particle information used (e.g., positions, velocities, halo membership at previous snapshots), the method for assigning particles in major mergers, and the method for deciding which of two overlapping haloes is the host halo in major mergers.  Names following the halo finders are the co-authors who ran the halo finders for this study and provided the following descriptions, who are not always the same as the original halo finder authors.

\subsection{Common Terms}

Throughout this paper, $\rho_c= {3H^2}/{(8\pi G)}$ refers to the critical density.  We use $R_{YYYc}$ to indicate the radius from a halo centre within which the average enclosed density is $YYY \times \rho_c$ (including substructure).  Similarly, $M_{YYYc}$ refers to the total mass enclosed within $R_{YYYc}$.  Some halo finders also use $R_\mathrm{vir}$, corresponding to an average enclosed density $\rho_\mathrm{vir}$ as defined in \cite{mvir_conv}.  The term ``$\vmax$'' refers to the maximum circular velocity; i.e., the maximum value of $\sqrt{G M(<R) / R}$ over a halo's radial mass profile.  Finally, ``position-space'' information refers to particle positions, ``velocity-space'' information refers to particle velocities, ``phase-space'' information refers to both particle positions and velocities, and ``temporal'' information refers to the evolution of particles' halo memberships over time.

\subsection{\ahf\ (Knebe)}
The halo finder
\ahf\footnote{\ahf\ is freely available from
  \url{http://www.popia.ft.uam.es/AMIGA}} \citep[\textsc{AMIGA}
Halo Finder,][]{AHF}, is an improvement of the \textsc{MHF}
halo finder \citep{Gill04}, which employs a recursively refined grid
to locate local overdensities in the density field. The identified
density peaks are then treated as centres of prospective haloes. The
resulting grid hierarchy is further utilized to generate a halo tree
readily containing the information which halo is a (prospective) host
and subhalo, respectively. Halo properties are calculated
based on the list of particles asserted to be gravitationally bound to
the respective density peak. To generate this list of particles we
employ an iterative procedure starting from an initial guess of
particles. This initial guess is based upon the distance of each prospective centre to its nearest more massive (sub-)halo where all particles within a
sphere of radius half this distance are considered prospective (sub-)halo
constituents. This tentative particle list is then used in an
iterative procedure to remove unbound particles and the final particle
list is truncated at some user pre-defined overdensity criterion.

The tree for each halo consists of one trunk and several branches where the trunk is the continuation of the main host halo and the branches represent the subhaloes \citep[see Fig.\ 1 in][]{AHF}. While there are various options in \ahf\ to pick the trunk, the default mode (also applied here) is to recursively follow the branch containing the most particles. This choice certainly leaves its imprint during major merger events studied here.

\subsection{\hbt\ (Han)}

\hbt\ \citep[Hierarchical Bound-Tracing algorithm;][]{HBT} is a tracking (sub)halo finder. Isolated haloes are first identified with a standard Friends-of-Friends (FoF) algorithm \citep{Davis85}. Within each isolated halo, the self-bound part is defined as a central subhalo. Starting from the highest redshift, subhaloes are then tracked down to later snapshots to link to their descendent haloes, by finding host haloes for the progenitor particles.  When two or more subhaloes are linked to a common descendent halo,  we compare the current self-bound mass of the progenitor subhaloes, and define their self-bound remnants, except the most massive remnant, as satellite subhaloes. The current central subhalo is re-defined to be the self-bound part out of all the particles in the host excluding satellite particles, while its progenitor is defined as the one that produced the most massive remnant. The tracking process is then continued for all the subhaloes including central and satellites down to the final output of the simulation.  The position and velocity for HBT subhaloes are defined using the 25\% of particles with the lowest local potential energy.

\subsection{\rockstar\ (Behroozi \& Mao)}

The \rockstar{} halo finder\footnote{\url{https://bitbucket.org/gfcstanford/rockstar}} \citep{Rockstar} adaptively shrinks phase-space isodensity contours to identify peaks in phase-space density.  Particles within an isodensity contour that contains only one peak are grouped into a single halo (or subhalo); the halo's position and velocity are average values for particles near the phase-space peak (typically, within $0.1 \rvir$).  When an isodensity contour contains multiple peaks, particles are assigned to the closest halo in phase space, determined by the metric $d(h,p)$:
\begin{eqnarray}
\label{e:membership}
d(h, p) & = & \left(\frac{|\vec{x}_h-\vec{x}_p|^2}{r_{\mathrm{dyn,vir}}^2} + \frac{|\vec{v}_h-\vec{v}_p|^2}{\sigma_v^2}\right)^{1/2}\\
r_{\mathrm{dyn,vir}} & = & \vmax t_\mathrm{dyn,vir} = \frac{\vmax}{\sqrt{\frac{4}{3}\pi G \rho_{vir}}} 
\end{eqnarray}
where $h$ is the halo, $p$ is the particle, $\sigma_v$ is the halo's current velocity dispersion, $\vmax$ is its current maximum circular velocity, and $\rho_{vir}$ is the virial overdensity from \cite{mvir_conv}.  Because particles are assigned to haloes before the final masses of the haloes are known, using $\vmax$ and $\sigma_v$ (which are both consistently measured even deep inside the halo potential well) improves particle assignment stability.  When two haloes overlap, the halo with the larger number of assigned particles is generally assumed to be the host halo.  However, in cases where two haloes are within a factor of 0.6 in $\vmax$, information on which halo was the host halo at the previous timestep is used to determine which halo will be labelled the host halo at the current timestep.

\subsection{\subfind\ (Muldrew \& Srisawat)}

\subfind\ \citep{Springel01} identifies gravitationally bound, locally overdense regions in a halo.  Initially, a FoF finder with linking length $b$ is used to identify haloes to be processed by \subfind. The density of the particles within these haloes is then estimated in an SPH-like (Smoothed Particle Hydrodynamics) fashion using an adaptive kernel interpolation with $N_{\rm dens}$ neighbours within the full volume.  Locally overdense regions are identified by considering each particle in order of density and searching for saddle points using the $N_{\rm ngb}$ nearest neighbours.  Particles with a higher density than their neighbours are used to define new candidate subhaloes.  Particles with neighbours that are of higher density, and are attached to a single substructure, become members of that substructure.  Finally, particles with denser neighbours that are attached to two different substructures are considered saddle points.  These candidate subhaloes are then iteratively tested for self-boundedness.  Starting with the lowest-density saddle point, a hierarchy of substructure is determined.  Subhaloes are defined as self-bound structures enclosed within an isodensity contour passing through the saddle point and containing a minimum of $N_{\rm ngb}$ particles.  Particles that are not assigned to any substructure are added to the `background halo.'  This is the largest subhalo that was found in the FoF halo, which is also tested for self-boundedness.  For this study we used the parameters $b=0.2$ and $N_{\rm dens}=N_{\rm ngb}=20$.  A more detailed description of \subfind\ can be found in \S{}4.2 of \cite{Springel01}.

\begin{figure}
\plotgrace{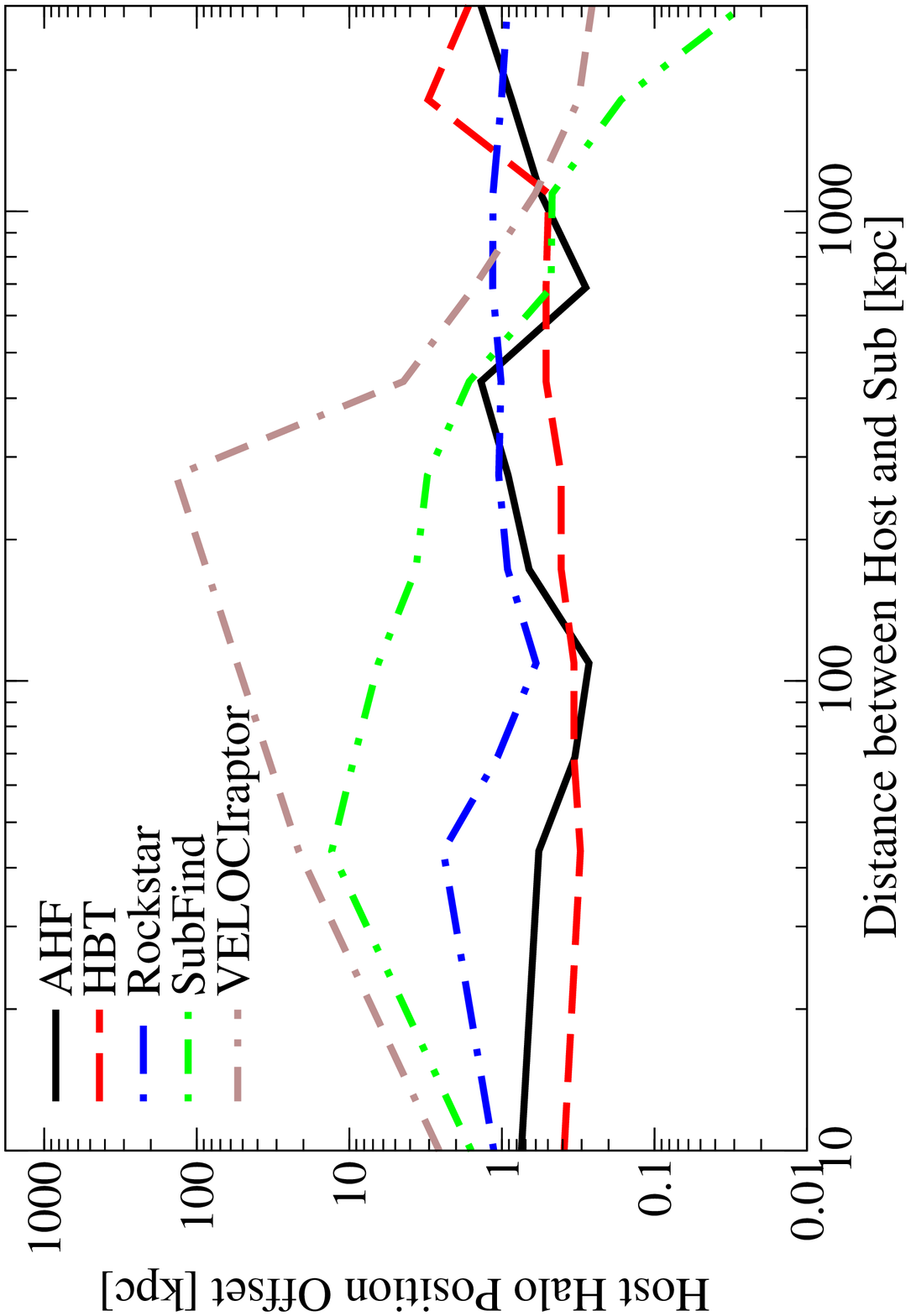}\\[-5ex]
\plotgrace{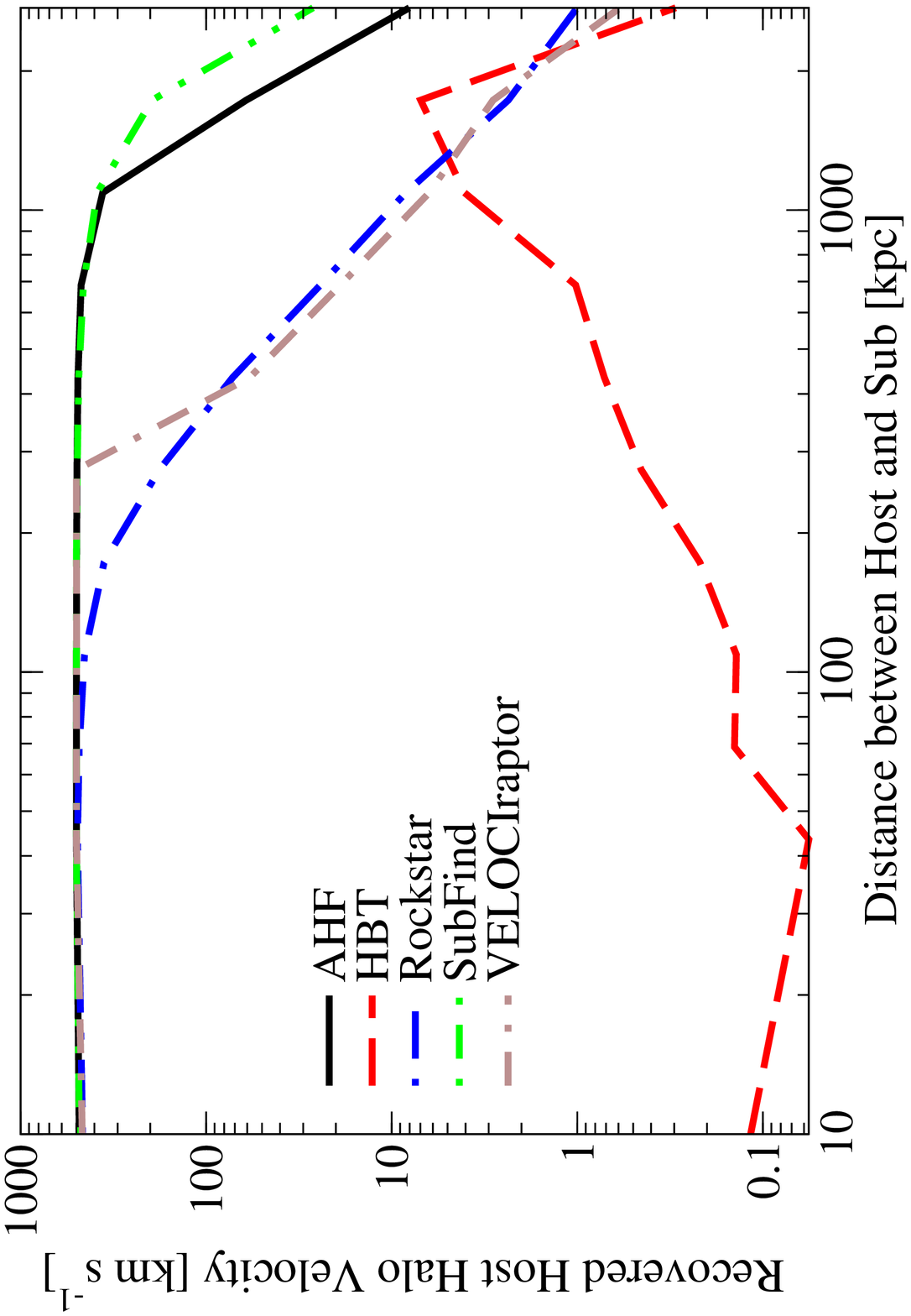}\\[-5ex]
\plotgrace{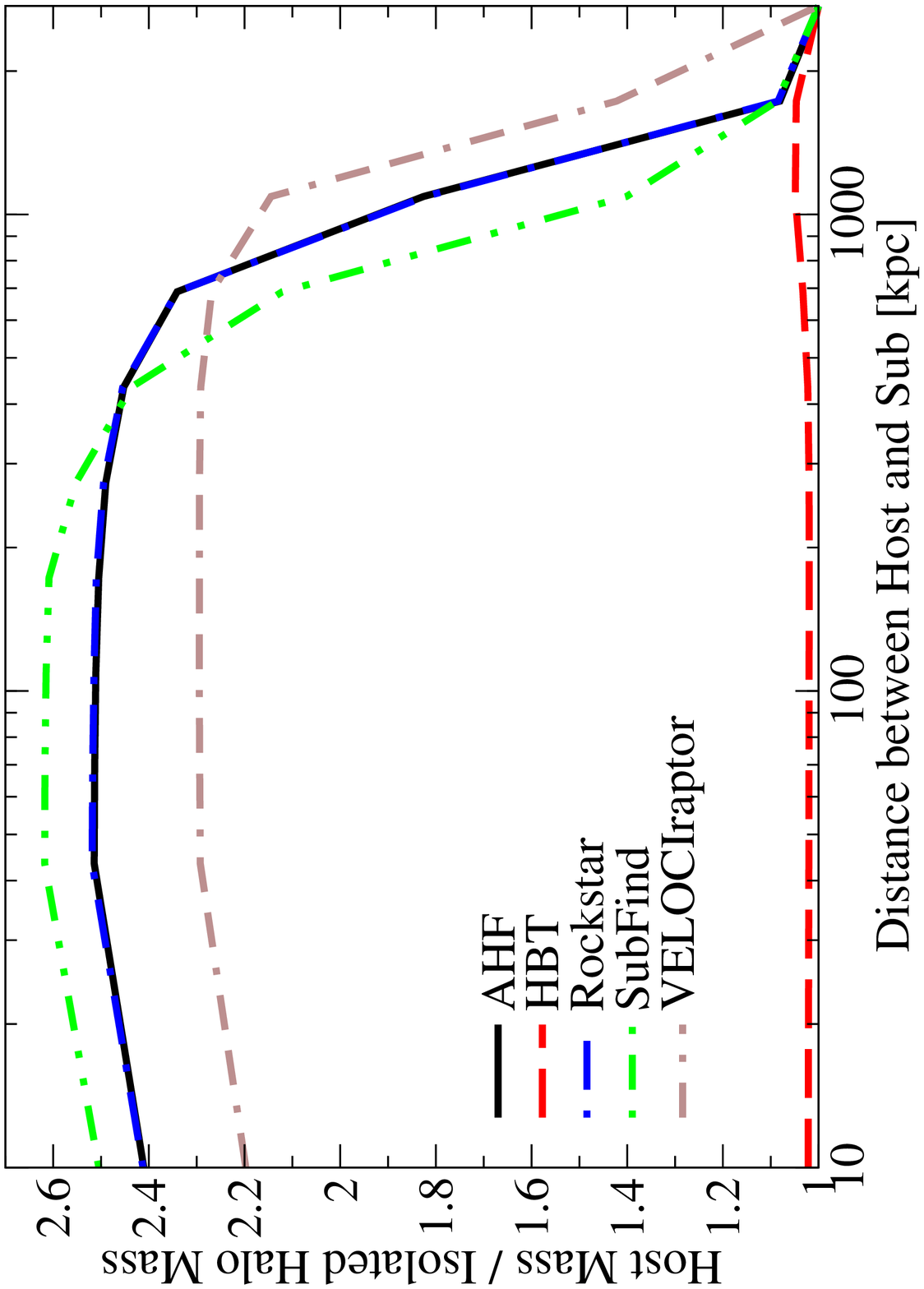}
\caption{Tests with overlapping identical mock halo profiles (\S \ref{s:static}) with a 1000 km s$^{-1}$ velocity offset, as a function of the distance between the halo centres.  In these tests, the ``host halo'' is taken to be the larger of the two returned haloes from each halo finder.  \textbf{Top} panel: offset between the input and recovered position of the host halo profile.  \textbf{Middle} panel: recovered host halo velocity; note that most of the halo finders here include substructure when calculating the host velocity.  The average velocity of all particles when the two halo profiles overlap completely is 500 km s$^{-1}$.  \textbf{Bottom} panel: Recovered host halo mass compared to the recovered halo mass for the largest separation of the two mock haloes (``Isolated halo mass'').}
\label{f:static_host}
\end{figure}

\subsection{\velociraptor\ (Elahi)}
\velociraptor\footnote{\url{https://bitbucket.org/pelahi/velociraptor-stf/}} \citep[a.k.a. {\sc STructure Finder} or {\sc STF}]{Elahi11} is a (sub)halo finder that identifies objects in a two-step process. First, haloes are identified using a FoF algorithm, where candidate haloes identified by a 3DFoF algorithm are pruned of any artificial particle bridges using a 6DFoF and the velocity dispersion of the FoF group. The 6DFoF is also used to flag major mergers, that is the presence of two (or more) large phase-space dense cores in the FoF halo. Here we follow the normal convention and treat the smaller object(s) as a subhalo and the larger as a host halo. These field objects are then searched for substructures by identifying particles that appear to be dynamically distinct from the mean halo background, i.e., particles which have a local velocity distribution that differs significantly from the averaged background halo. These dynamically distinct particles are linked with a phase-space FoF algorithm into substructures. Since this approach is capable of not only finding subhaloes, but the unbound tidal debris surrounding them as well as tidal streams from completely disrupted subhaloes, for this analysis we also ensure that a group is self-bound. 

In similar-mass mergers, the mean field is an equal combination of both haloes, thus neither core will contain (many) particles that appear locally dynamically distinct. Hence once the cores overlap enough in phase-space, the system will no longer be flagged as a merger and {\em the smaller core will not necessarily appear as a dynamically distinct substructure either}.

\section{Static Mock Profile Tests}

\label{s:static}

Because no common definition exists for the correct properties of cosmologically-simulated haloes \citep{Knebe13}, synthetically-generated haloes are one of the few ways to test halo finder accuracy.  We adopt the spherical mock host halo NFW profile described and tested in \cite{Muldrew11}, which has a mass\footnote{In this case, the critical density is calculated for a flat $\Lambda$CDM cosmology with $\Omega_M = 0.3$, $h=0.73$, at $z=0$.} of $M_{200c} = 1.04\times 10^{14} M_\odot$ (other parameters include $R_{200c} = 944.0$ kpc, $R_s = 259.6$ kpc, $\vmax = 715$ km s$^{-1}$).  The halo profile extends to 2.75 $R_{200c}$ and is sampled with $\sim$ 1.5 million particles, each of mass $1.37\times10^{8}\Msun$. 

We place this halo in the centre of an empty volume and place a duplicate copy of the halo with a velocity offset of 1000 km s$^{-1}$ and a distance offset between 0 and 2700 kpc.  The chosen velocity offset is typical for cosmological major mergers on first infall \citep{BehrooziTree}; as the haloes' velocity dispersions are both $\sigma_v = 754$ km s$^{-1}$, this places their centres at an offset of $1.33\sigma_v$ in velocity space.  Since the profiles are identical, there is no ``correct'' host halo or subhalo choice; however, halo finders typically tag overlapping/merging haloes as ``host'' and ``subhalo,'' respectively. Note that this choice might be arbitrary, but we also decided to follow this notion by referring to the \textbf{larger} of the two recovered objects as the ``\textbf{host halo}'' and the \textbf{smaller} as the ``\textbf{subhalo}.''
Properties for this ``host halo'' are therefore calculated on the combined profiles, whereas properties of the ``subhalo'' are (ideally) calculated on particles from only one of the individual profiles.

We show results for recovery of the centre, velocity, and mass for the host halo in Fig.\ \ref{f:static_host}.  The upper-most panel compares the input position to the one returned for the halo tagged as ``host'' by the respective finder. The positional offset is typically always smaller than 10 kpc.  However, \velociraptor\ calculates halo centres as the centre-of-mass of the innermost 10\% of particles within the halo radius (determined iteratively), which leads to larger offsets when the two mock profiles are within 300 kpc.  Some differences are also notable in the velocity calculations -- as presented in the middle panel.  \ahf\ and \subfind\ report averaged particle velocities within the full host halo radius, including particles from the subhalo. However, \rockstar\ uses a velocity measured closer to the halo centre (at 0.1 $\rvir$) and therefore averages fewer particles from the subhalo when determining the host halo's velocity, until the subhalo approaches much closer to the host halo's centre.  \velociraptor\ similarly attempts to exclude most substructure when calculating host halo velocities. \hbt\ does not include substructure at all when calculating the host halo velocity, and so the recovered velocity is relatively independent of the subhalo's position. The same explanation holds for \hbt\ in the bottom panel, where the recovered mass is compared against the input host mass.  We note that for \hbt{}, which requires a sequence of snapshots in order to recover subhaloes, the haloes were processed (and particles tracked) in order of largest (2700 kpc) to smallest (0 kpc) distance separations.

\begin{figure*}
\plotgrace{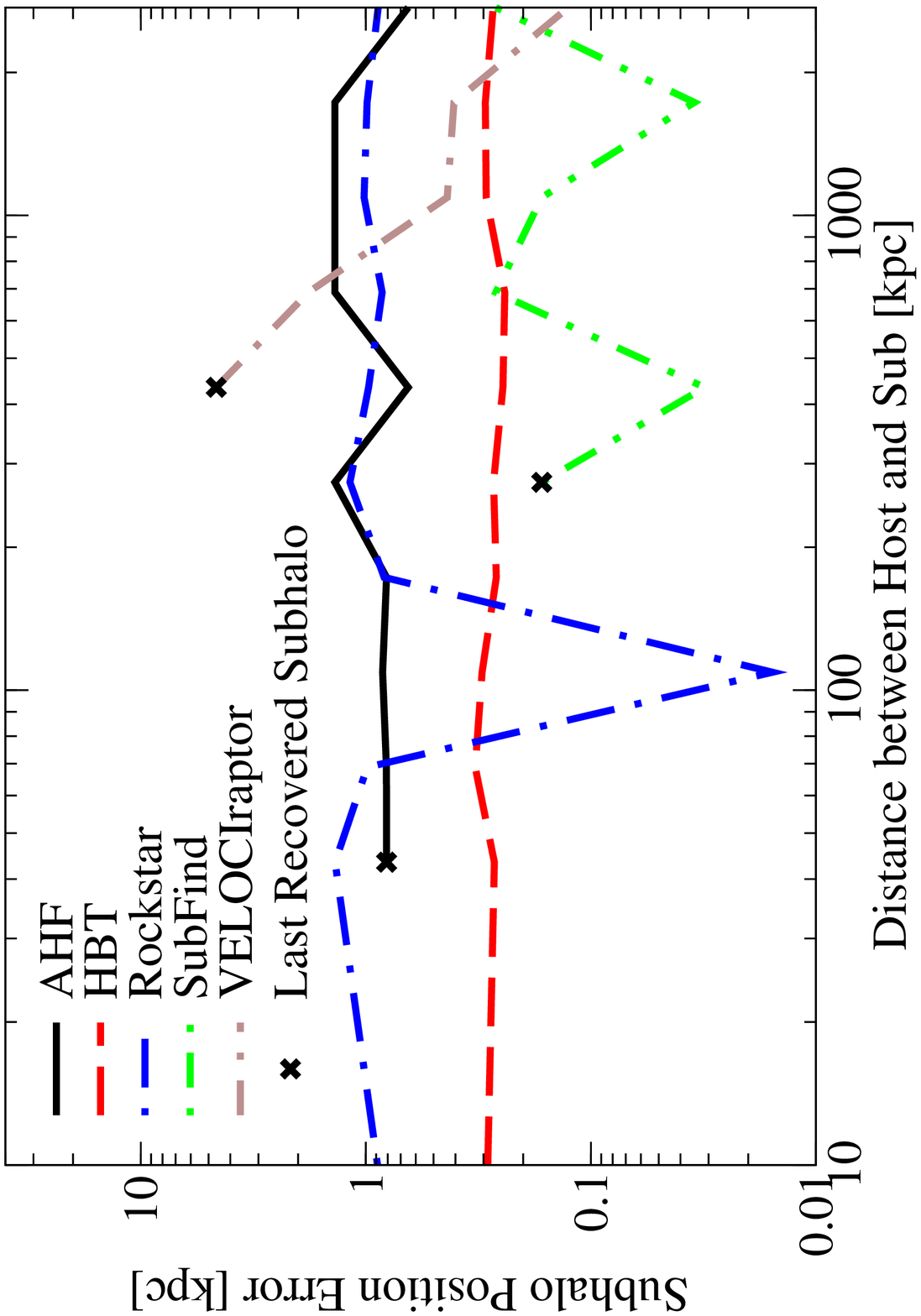}\plotgrace{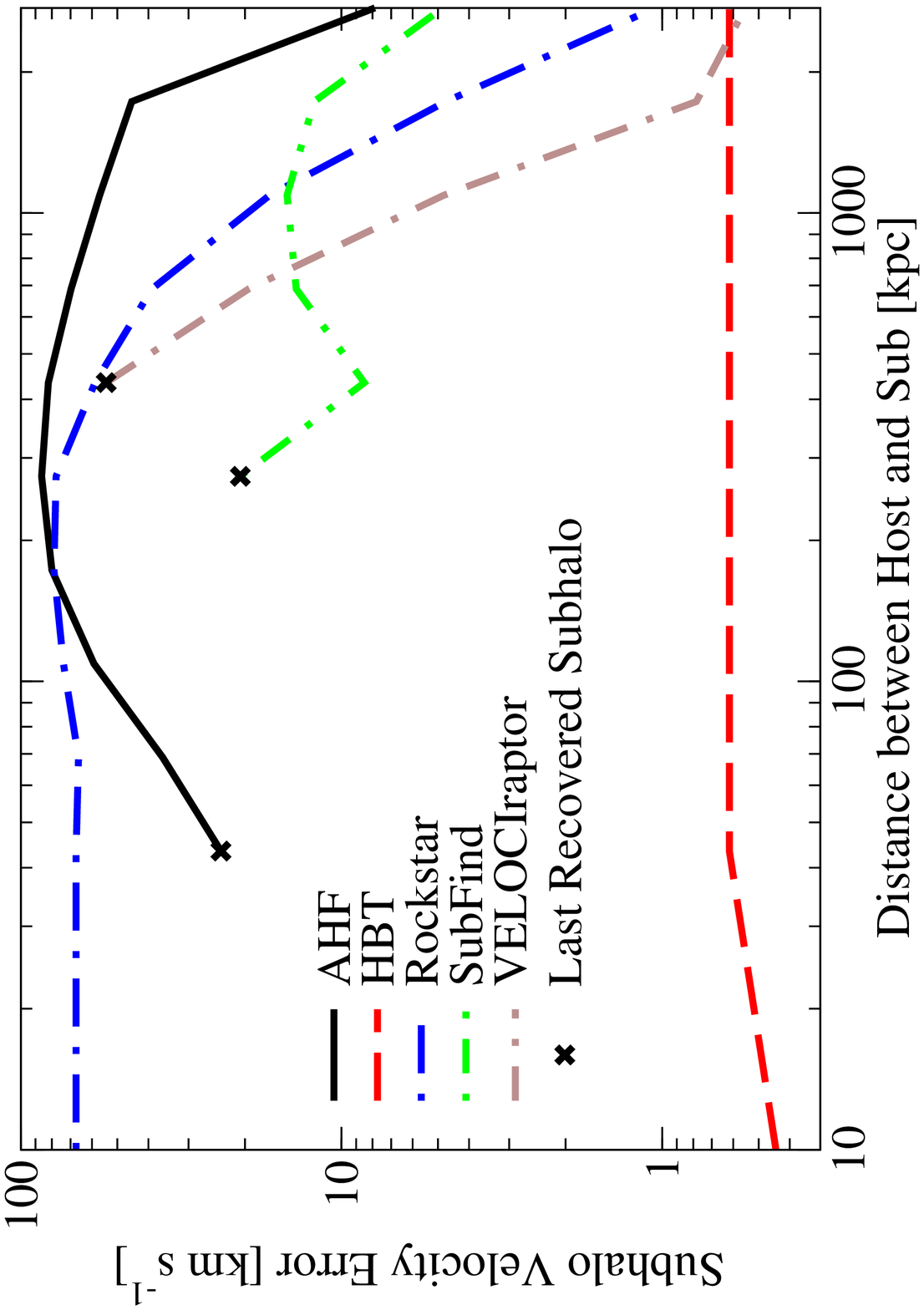}\\[-5ex]
\plotgrace{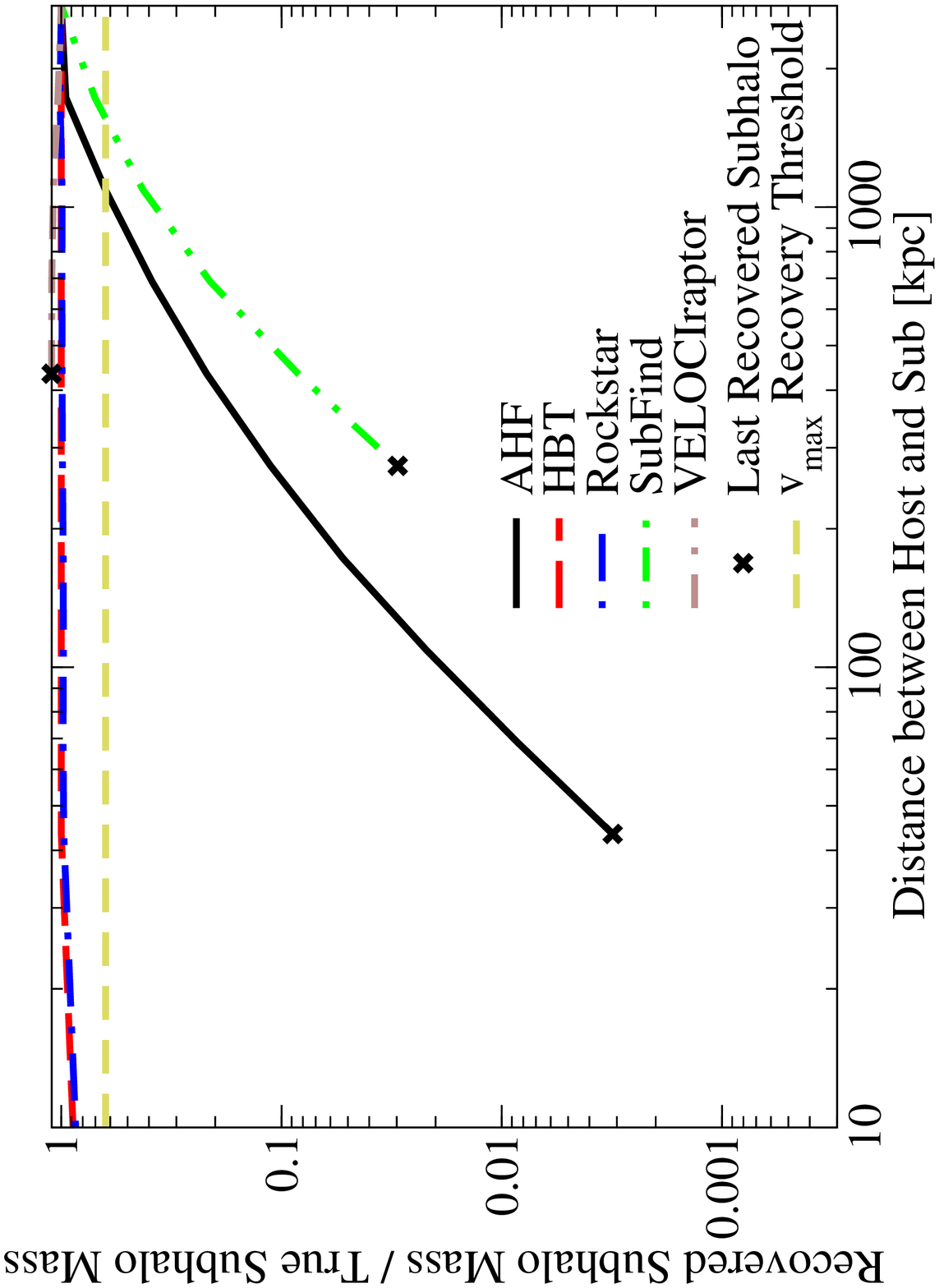}\plotgrace{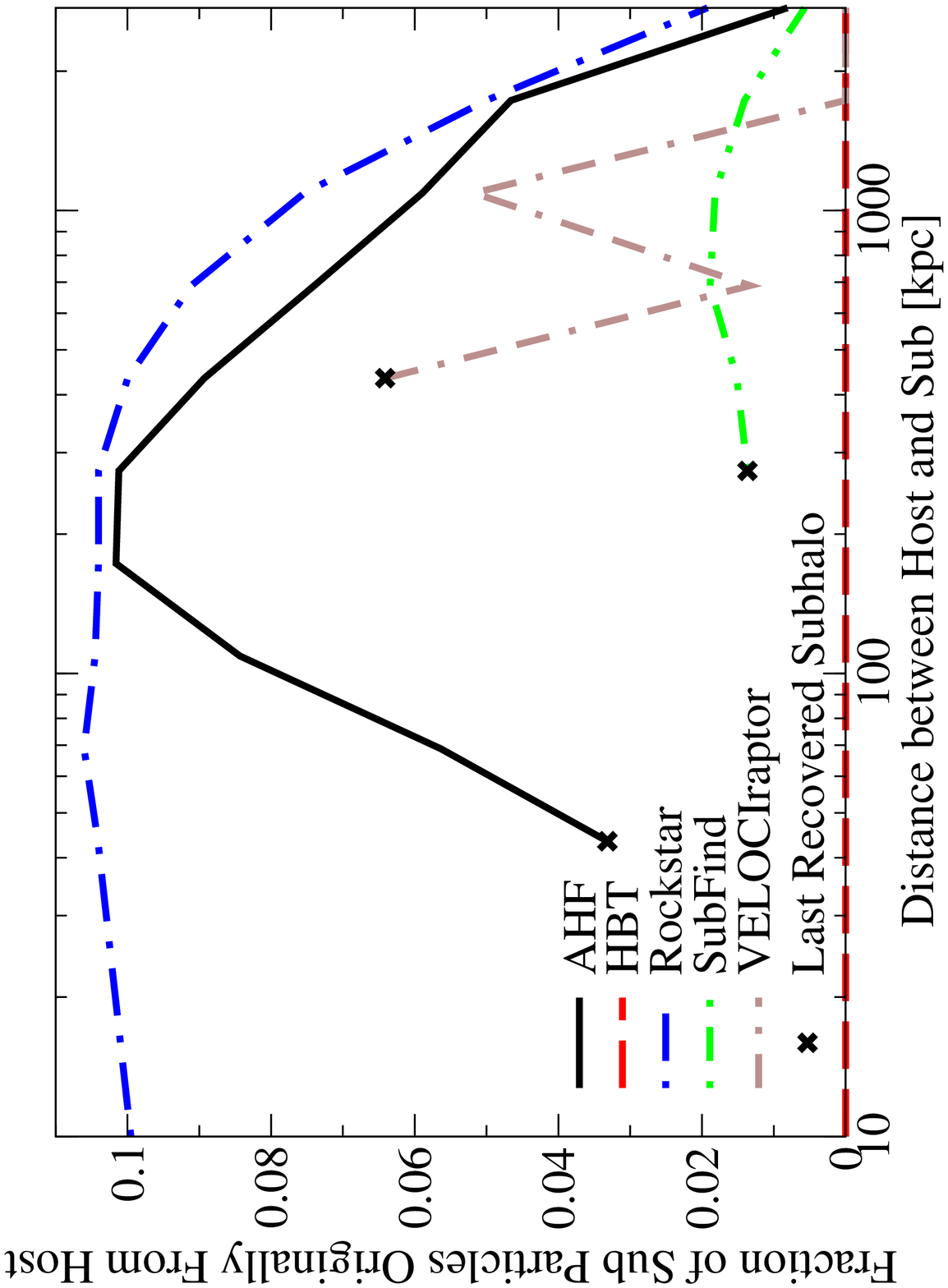}
\caption{Tests of overlapping identical mock halo profiles (\S \ref{s:static}) as a function of the distance between the halo centres.  In these tests, the ``subhalo'' is taken to be the smaller of the two returned haloes from each halo finder.  \textbf{Top-left} panel: errors in recovering the subhalo's position.  \textbf{Top-right} panel: errors in recovering the subhalo's velocity.  \textbf{Bottom-left} panel: ratio of the recovered subhalo mass to the input subhalo mass.  The \textit{amber dashed line} shows the mass threshold below which it is impossible to recover the true $\vmax$ of the halo (see \S \ref{s:static}); see, however, Fig.\ \ref{f:vmax_loss}.  \textbf{Bottom-right} panel: fractional number of particles assigned to the subhalo which were originally from the input host halo's profile.}
\label{f:static}
\end{figure*}

Fig.\ \ref{f:static} shows recovery of the position, velocity, and mass of the subhalo, as well as the fraction of contamination from the host halo's particles.  As above, comparison is made to the input positions and velocities of the halo profile closest to the halo finder's recovered subhalo centre.  Extremely good agreement (within $\sim$ 1 kpc) is seen for position recovery, and the velocity recovery is generally within 10\% of the relative host and subhalo velocities.  Notably, no advantage in accuracy is seen for phase-space algorithms.  When the haloes are highly overlapping, particle membership is determined largely by a velocity cut; yet, since the halo velocity distributions \textit{also} overlap, this results in an asymmetric truncation of the subhalo's velocity distribution.  Without a special averaging technique, this leads to a systematic positive radial velocity bias.  The situation is exactly reversed for the position-space algorithms: particles within a given radial aperture of the subhalo centre will be contaminated with host particles, leading to a systematic \textit{negative} radial velocity bias.  That said, this effect is small in magnitude even for the worst-case scenario shown here.  Naturally, the effect does not exist at all for HBT, as particle contamination from the host is not an issue.

\begin{figure}
\plotgrace{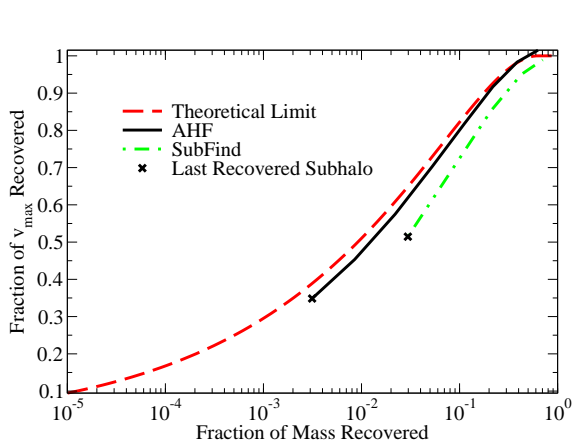}
\caption{A comparison of fractional mass loss to fractional $\vmax$ loss in the recovered mock subhalo properties, compared to the theoretical expectation from the NFW profile used.}
\label{f:vmax_loss}
\end{figure}

In terms of mass recovery, the position-space finders are at a severe disadvantage.  Ordinarily, position-space finders can collect particles at large radii around substructure density peaks and take advantage of the fact that the high-velocity particles belonging to the host halo will be removed in the gravitational unbinding stage.  In effect, this performs a quasi-phase-space particle selection.  However, since the velocity dispersion of the two test profiles is the same, this technique no longer works.  The position-space finders therefore end up truncating the mass profile of the subhalo as soon as the profiles begin to overlap (Fig.\ \ref{f:static}).  $\vmax$ recovery tends to be much better: given an NFW profile which has been spherically truncated, $\vmax$ is very insensitive to the amount of truncation, as shown in Fig.\ \ref{f:vmax_loss}.  Yet, $\vmax$ recovery is impossible if the halo itself cannot be found, which happens below 300 kpc for \subfind\ and 45 kpc for \ahf.  These problems for position-space finders are also seen, albeit to a lesser extent, for minor mergers \citep{Muldrew11,Knebe11}.  The phase-space finders do very well by comparison, with \rockstar\ and \velociraptor\ recovering between 90-95\% of the original halo mass.  When two haloes overlap significantly in phase space, \velociraptor\ treats them as a single halo; however, \rockstar\ continues to recover two haloes as long as the haloes' innermost density peaks are distinguishable.  While both \velociraptor\ and \rockstar\ are phase-space algorithms, this difference allows \rockstar\ to recover properties of haloes in major mergers at much closer separations.  As expected, \hbt\ is able to perfectly recover the input halo's mass.

Finally, we address the issue of subhalo purity (lower right panel of Fig.\ \ref{f:static}).  \rockstar's subhalo is contaminated at up to the 10\% level by particles from the host halo, which is in agreement with the expected fraction of host particles which are $1.33\sigma_v$ offset from the host halo's central velocity.  For \ahf\ and \subfind, the level of contamination depends on how aggressively they try to recover the subhalo's radial profile.  \subfind\ is more conservative (and purer), with the result that it recovers much less mass as compared to \ahf.  By definition, \hbt\ does not have any purity issues; particles initially assigned to the host are never allowed to be assigned to the subhalo.

\section{Dynamic Mock Infall Tests}

\label{s:dynamic}

\begin{figure}
\plotgrace{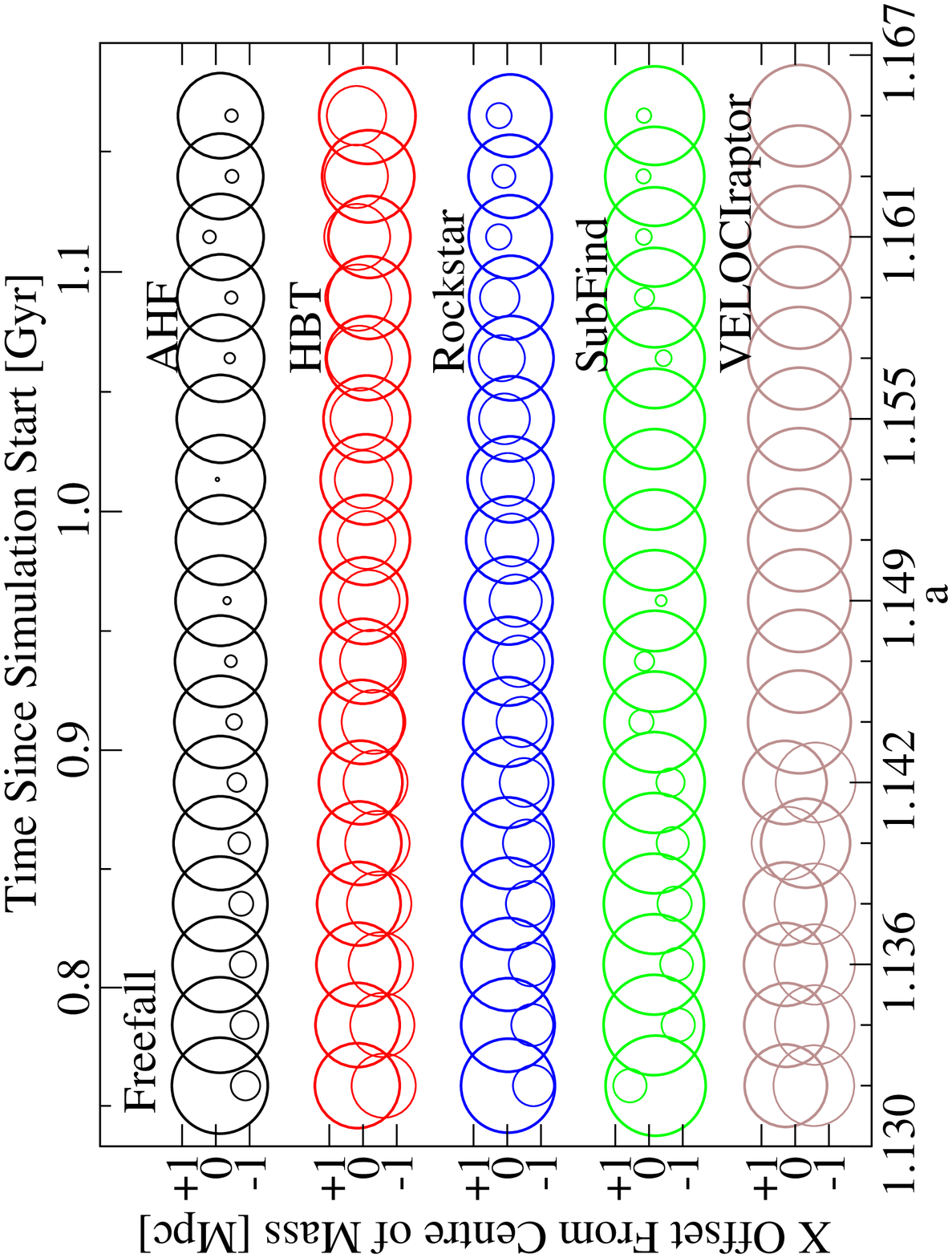}\\[-5ex]
\plotgrace{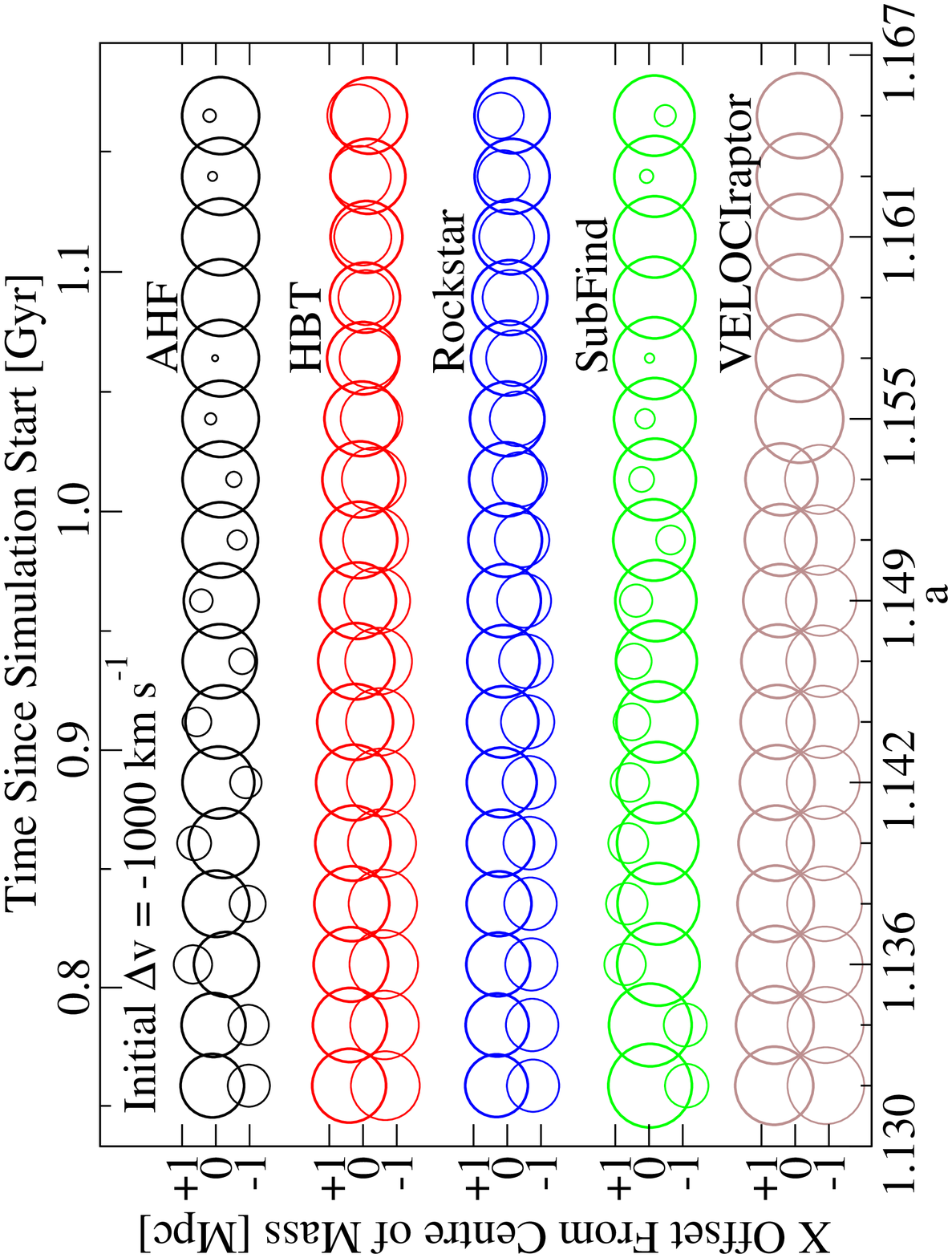}
\caption{Time series plots for dynamically simulated twin mock haloes; the haloes were released from rest at a separation of 2700 kpc in the \textbf{top} panel, and given a more realistic initial velocity offset of 1000 km s$^{-1}$ in the \textbf{bottom} panel (see \S \ref{s:dynamic}).  Each circle corresponds to a halo at the given snapshot; circle radii correspond to the halo radius---i.e., they are proportional to the cube root of the halo mass---and the \textbf{bold} circle corresponds to the host halo.  The snapshots are equally spaced in scale factor from $a=1.132$ to $a=1.165$, about 420 Myr, and cover a close interaction between the two haloes ($a=1.15$ and $a=1.16$ for the freefall and velocity offset cases, respectively).  Each tick mark on the bottom axis corresponds to a separate snapshot.  Results from each halo finder have been spatially offset for clarity.}
\label{f:dynamic}
\end{figure}

We next consider a more realistic, dynamically simulated test.  The initial conditions are two identical mock haloes (the same as described in \S \ref{s:static}) placed at an initial separation of 2700 kpc.  We consider two initial velocities for the haloes.  In the ``Freefall'' test, the haloes are released from rest.  In the more cosmologically realistic ``Dynamic'' test, the first halo begins at rest, and the second halo is given a 1000 km s$^{-1}$ velocity offset, aimed toward a point offset 140 kpc from the centre of the first halo.  These initial conditions (taken to occur at $a=1$) were simulated forward in time to $a=1.2$ by PKDGRAV2 \citep{Stadel09}, including background cosmological expansion according to a flat, $\Lambda$CDM cosmology with $\Omega_M = 0.3$ and $h=0.73$ at $z=0$; the assumed force resolution was 6.8 kpc.  Because an analytic solution for the merger of two haloes is not known, the main purpose of this test was to check the consistency over time of the returned halo properties.  

The positions and radii (with masses being proportional to the third power of the radii) of haloes in both tests are shown as a time-series plot in Fig.\ \ref{f:dynamic}.  As in the static tests, the position-space halo finders show marked artificial mass loss as the haloes approach each other.  In addition, these finders show a dramatic ``flip-flopping'' between which halo is assigned to be the host halo and which is assigned to be the subhalo.  To be fair, the extreme symmetry of the tests means that all halo finders which do not include some temporal information in deciding host and subhalo relationships will show similar behaviour.  However, this behaviour is also seen in cosmological simulations (\S \ref{s:cosmological}; see also \citealt{Tweed09,HBT,Rockstar,BehrooziTree,Srisawat13}).  Since \rockstar\ includes temporal information for host-subhalo assignments, it does not show flip-flopping; similarly, \hbt\ is immune.

One other feature is evident in the ``Freefall'' test.  In \rockstar's results, the subhalo appears to first grow ($a<1.155$) and then to lose mass again ($a>1.155$).  This arises because the cores of the merging haloes reach the centre of the potential well and begin to orbit rapidly, whereas the remainder of the material has a much slower orbital period.  When the subhalo core is in-phase with the velocity of the outer remnants, the recovered mass increases; when the subhalo core is out-of-phase, the remnant is assigned to the host halo \citep{Rockstar}.  This results in a periodic oscillation of the recovered halo masses, which is a fundamental limit of the phase-space algorithm employed.  This problem is avoided in \velociraptor\ because it no longer separates haloes once they begin to overlap significantly in phase space.  

\begin{figure}
\includegraphics[width=\columnwidth,type=eps,ext=.eps,read=.eps]{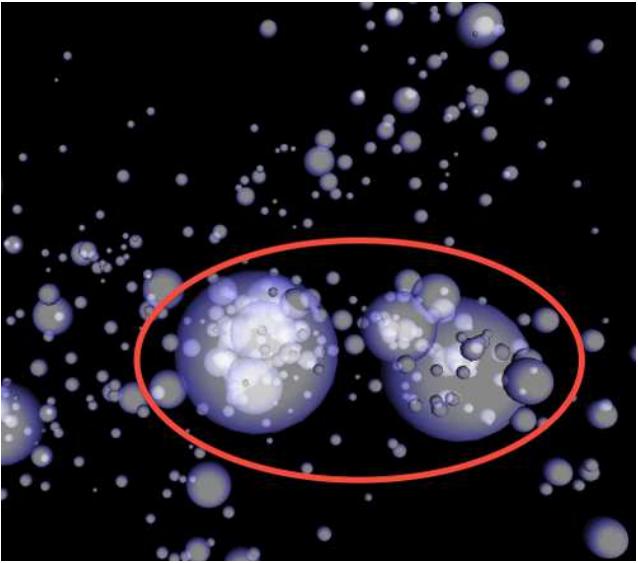}
\caption{A snapshot of the cosmological simulation described in \S \ref{s:cosmological}, showing the major merger we have selected for analysis (\textit{red ellipse}).  The spheres in this image correspond to the locations and radii ($R_{200c}$) of haloes returned by the \rockstar\ halo finder.  Full movies of the merger process for all halo finders are available online.\protect{\scriptsize{}$^{\ref{fn:url}}$}}
\label{f:snapshot}
\end{figure}

\begin{figure}
\vspace{-5ex}
\plotgrace{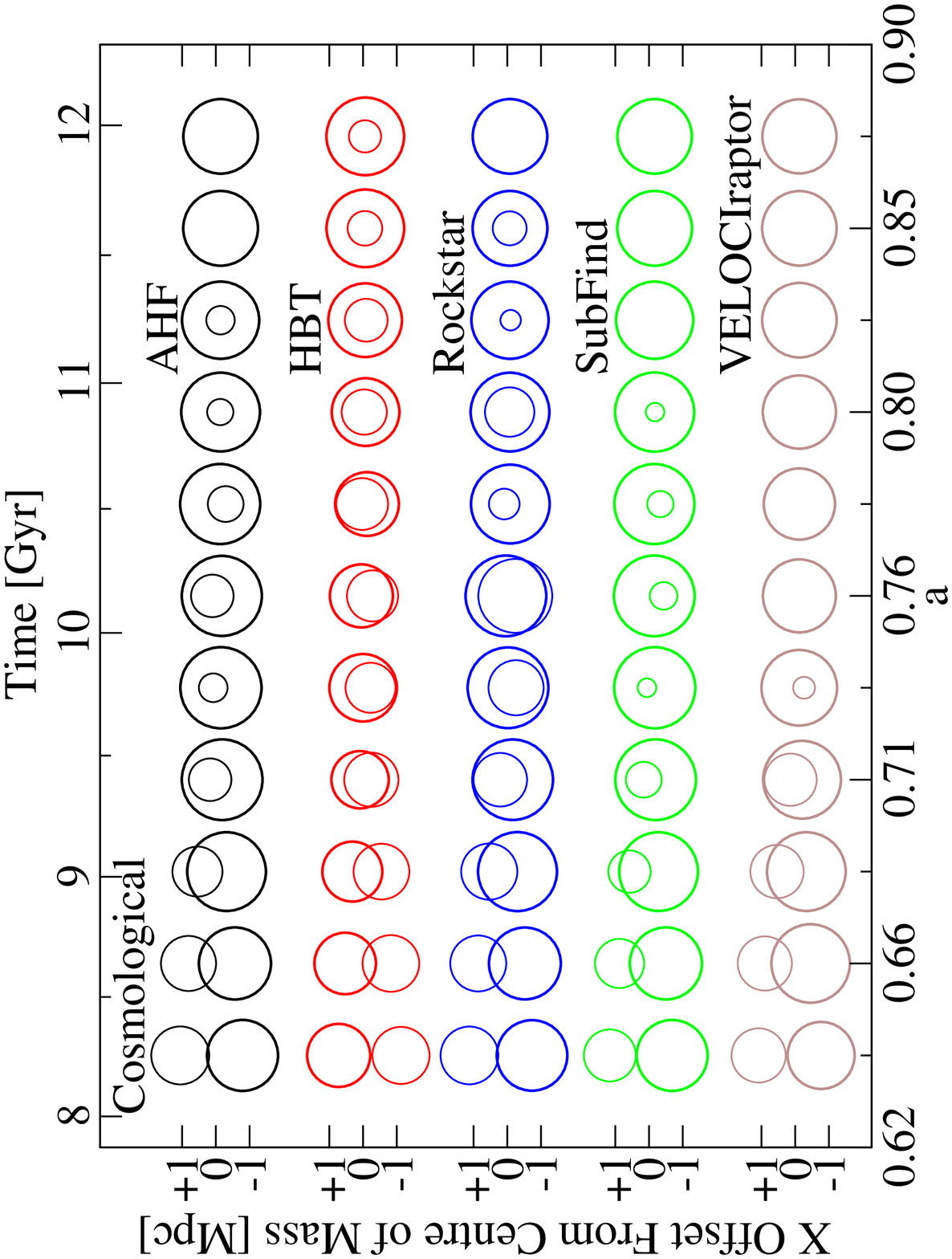}\\[-5ex]
\plotgrace{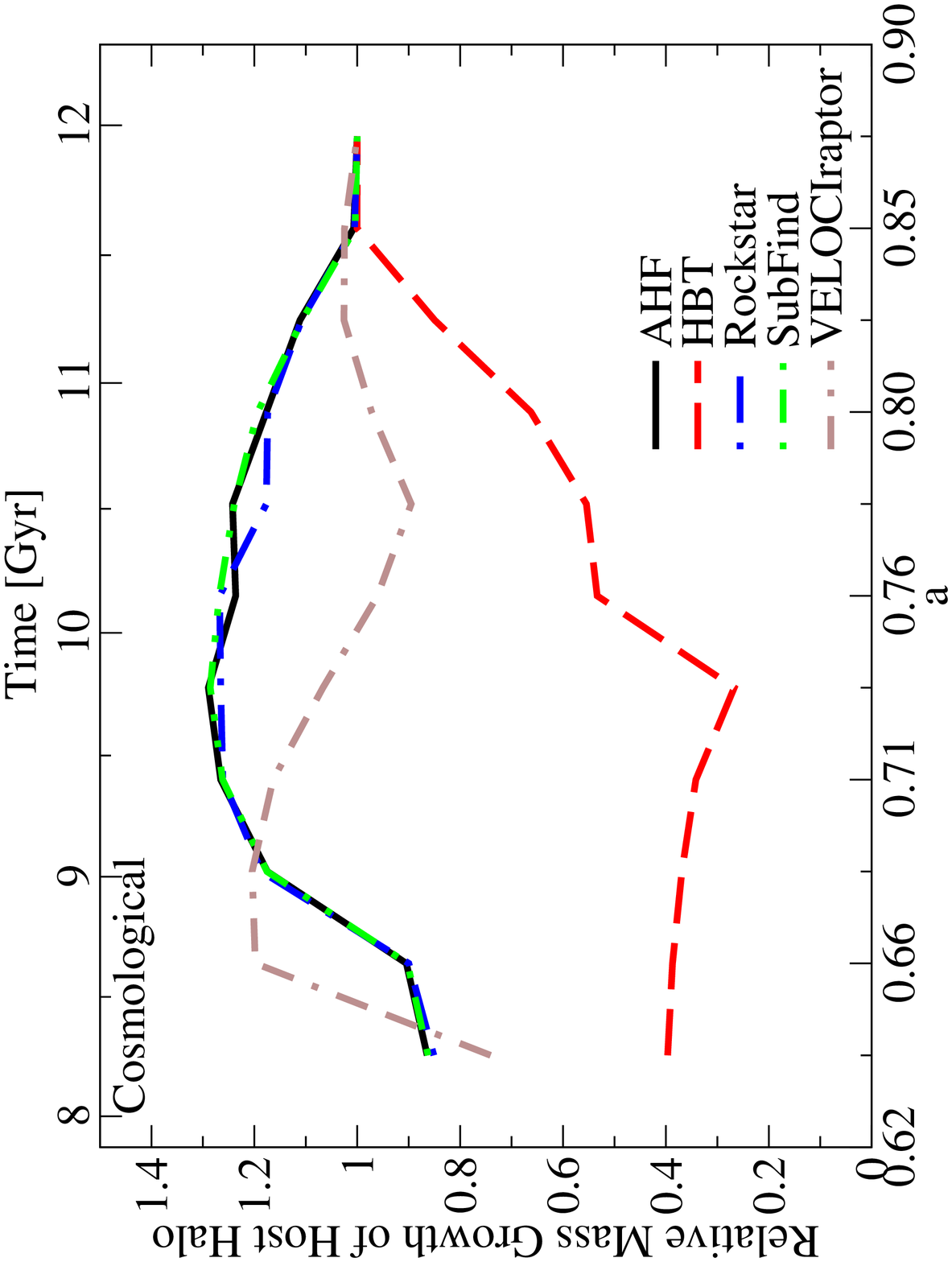}
\caption{Merger of the two largest haloes shown in Fig.\ \ref{f:snapshot}.  \textbf{Top} panel: time series plots of the merger, as returned by the different halo finders (as in Fig.\ \ref{f:dynamic}).  Each circle corresponds to a halo at the given snapshot; circle radii are proportional to the cube root of the halo mass, and the \textbf{bold} circle corresponds to the host halo.  Results from each halo finder have been spatially offset for clarity.  \textbf{Bottom} panel: mass growth history of the eventual host halo, normalized to the host halo mass at the last snapshot.  The snapshots are equally spaced in scale factor from $a=0.64$ to $a=0.87$, covering a range of 3.69 Gyr.  Each tick mark on the bottom axis corresponds to a separate snapshot.  }
\label{f:cosmological}
\end{figure}

\section{Cosmological Tests}

\label{s:cosmological}

\subsection{Simulation}

We make use of a simulation described in \cite{Srisawat13} and used for several studies emerging out of the Sussing Merger Trees comparison project. This simulation used the \textsc{GADGET-3} code \citep[an improved version of the code presented in][]{Springel05} to simulate 270$^3$ particles (mass resolution: $1.32\times 10^9 \Msun$) in a periodic box with side length 88.8 Mpc.  The adopted initial conditions were taken from the WMAP-7+BAO+$H_0$ best-fit \citep{WMAP7}; i.e., a flat, $\Lambda$CDM cosmology with parameters $\Omega_m = 0.272$, $\Omega_b = 0.0455$, $h = 0.704$, $n_s = 0.967$, and $\sigma_8 = 0.810$.  All the halo finders in this project analyzed this simulation as part of \cite{Perez13}.

\subsection{Individual Test Case}

\label{s:cosmo_individual}

We first examine a test case with an individual major merger, selected from the history of the second largest halo in the box at $z=0$.\footnote{The largest halo is still undergoing a major merger at $z=0$, so it is not possible to have a clean ``before'' and ``after'' comparison.}  A snapshot of the merger (mass ratio = 1:1.8) is shown in Fig.\ \ref{f:snapshot}, and movies of the halo catalogues returned by all finders are available online.\footnote{\label{fn:url}\url{http://slac.stanford.edu/~behroozi/MM_Movies/}} The returned haloes, pruned to exclude all but the host halo and the merging halo, are shown in the top panel of Fig.\ \ref{f:cosmological}.  The results of \S \ref{s:static} and \S \ref{s:dynamic} are instructive, as many of the same findings apply.  As in those sections, the position-space halo finders struggle to recover the masses of the subhalo, and show clear exchanges of host and subhalo relationships (e.g., \ahf\ at $a=0.78$, and \subfind\ at $a=0.76$).  Additionally, \rockstar\ shows substantial variation in the subhalo mass when it passes close to the centre; due to the coarser time resolution, this effect is dramatically exaggerated in comparison to the dynamic merger test in Fig.\ \ref{f:dynamic}, but the explanation is the same (see \S \ref{s:dynamic}).  Tree building algorithms can detect and repair this variability to some extent \citep[e.g.,][]{BehrooziTree}, regardless of the halo finder used.

\hbt, which had shown exemplary performance for the static and dynamic tests (\S \ref{s:static} and \S \ref{s:dynamic}), nonetheless has some issues with mass recovery in the cosmological test.  In Fig.\ \ref{f:cosmological}, it is clear at the beginning ($a=0.64$) that all halo finders except for \hbt\ find that the lower halo (corresponding to the leftmost halo in Fig.\ \ref{f:snapshot}) has a larger mass than the upper halo.  At this snapshot, the radii of the two merging haloes are barely touching---so it is not an issue of distinguishing between host and subhalo particle assignment.  Instead, haloes which are major mergers tend to accrete mass up to (and sometimes within) the virial radius of the larger halo \citep{BehrooziMergers}.  Because the corresponding friends-of-friends groups ``bridge'' well before the haloes come into contact \citep[see, e.g.,][]{Bolshoi}, \hbt\ limits the growth of the lower halo by allowing it only to consume particles belonging to its original FoF group.

\hbt\ also highlights an important issue with the definition of a halo.  Unlike the other halo finders, \hbt\ continues to track the smaller halo as a subhalo until $z=0$.  However, at this point, \hbt\ finds that the subhalo's position is within 9 kpc  of the host, on the same order as the force resolution of the simulation (7 kpc).  In addition, the velocity-space offset is only 100 km s$^{-1}$ from the host, which is very small compared to the velocity dispersion of the host (900 km s$^{-1}$).  Since the mass of \hbt's subhalo at $z=0$ is 20\% of the host mass, \textit{no algorithm} can robustly distinguish the subhalo particles from the host halo particles using phase space information alone.  While there is no doubt that \hbt's subhalo at $z=0$ is a self-bound structure, it is not clear that all applications would wish to treat it as separate from its host (see also discussion in \S \ref{s:discussion}), especially semi-analytical/abundance matching models and merger rate calculations.

In the lower panel of Fig.\ \ref{f:cosmological}, we show the mass growth of the host halo.  In mergers, the sum of the two progenitor haloes' masses can easily exceed the final halo mass after the merger; this is especially true in major mergers.  Major mergers dramatically raise the velocity dispersion in the resulting merged halo, with the result that many particles are unbound and that many of the remainder spend extended amounts of time orbiting beyond the halo radius \citep{Anderhalden11,BehrooziUnbound}.  As a consequence, the mass of the host increases by the mass of the smaller halo ($a=0.71$), but then rapidly falls as the merger raises the velocity dispersion.  At the end ($a=0.87$), the host mass only increases by 15\% compared to its original value.  \hbt\ again is the exception here as the two haloes remain clearly separated until $a=0.80$, which is when the two centres remain at the same position (Fig.\ \ref{f:cosmological}, upper panel). The subhalo then starts to transfer its mass to the host, and the end-mass of the merger approaches the result of the other finders.

\begin{figure*}
\vspace{-5ex}
\plotgrace{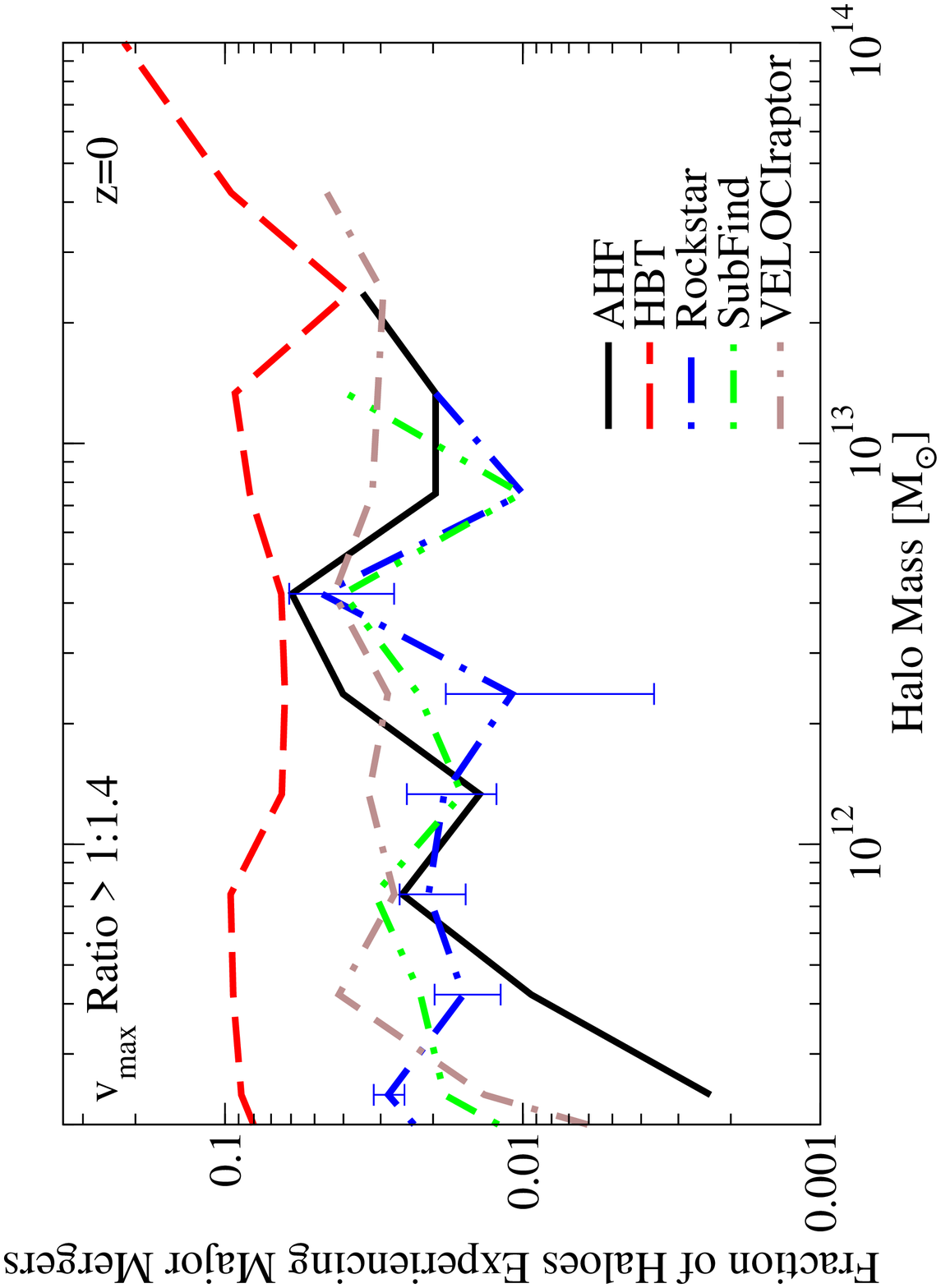}\plotgrace{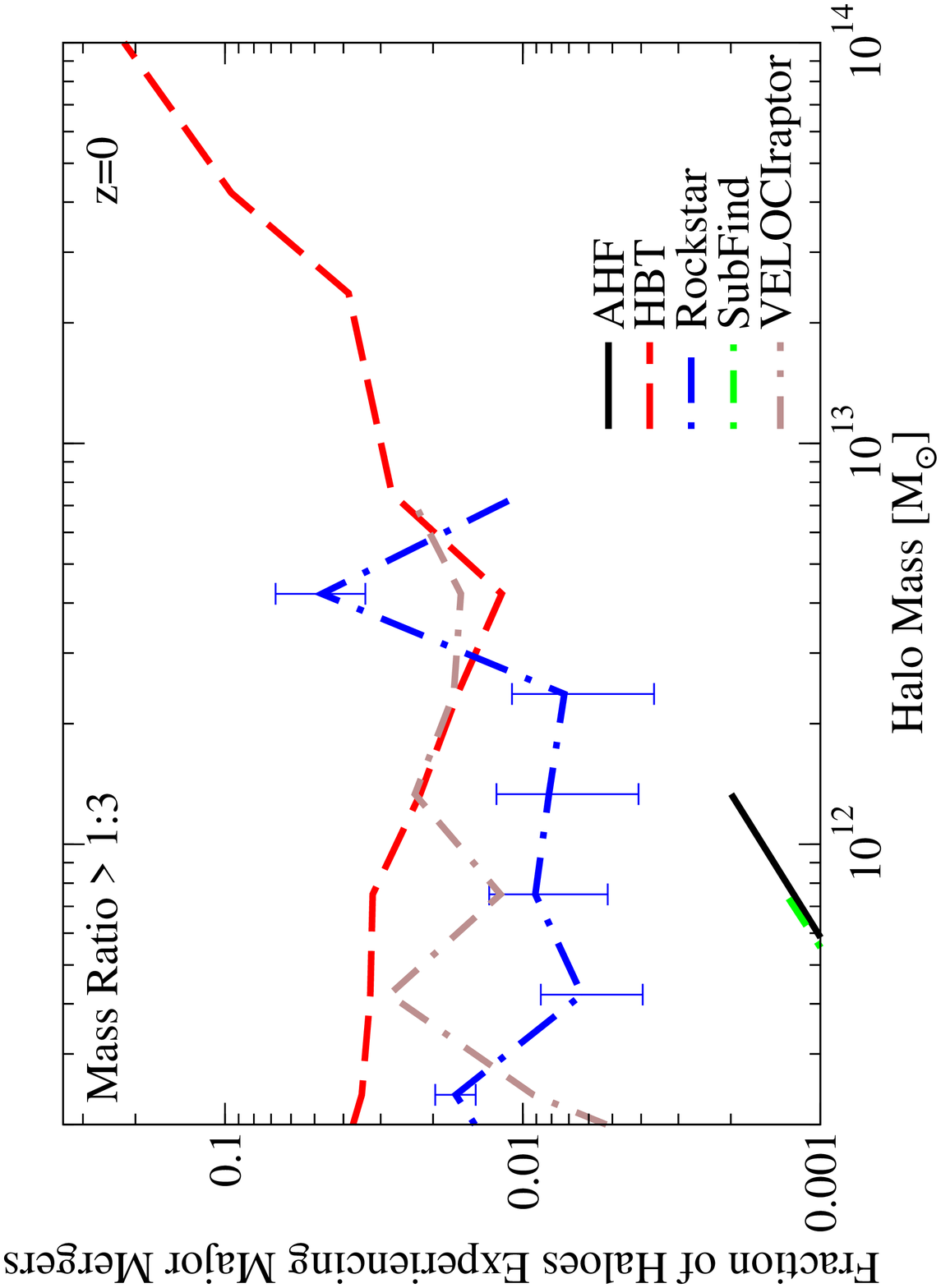}\\[-5ex]
\plotgrace{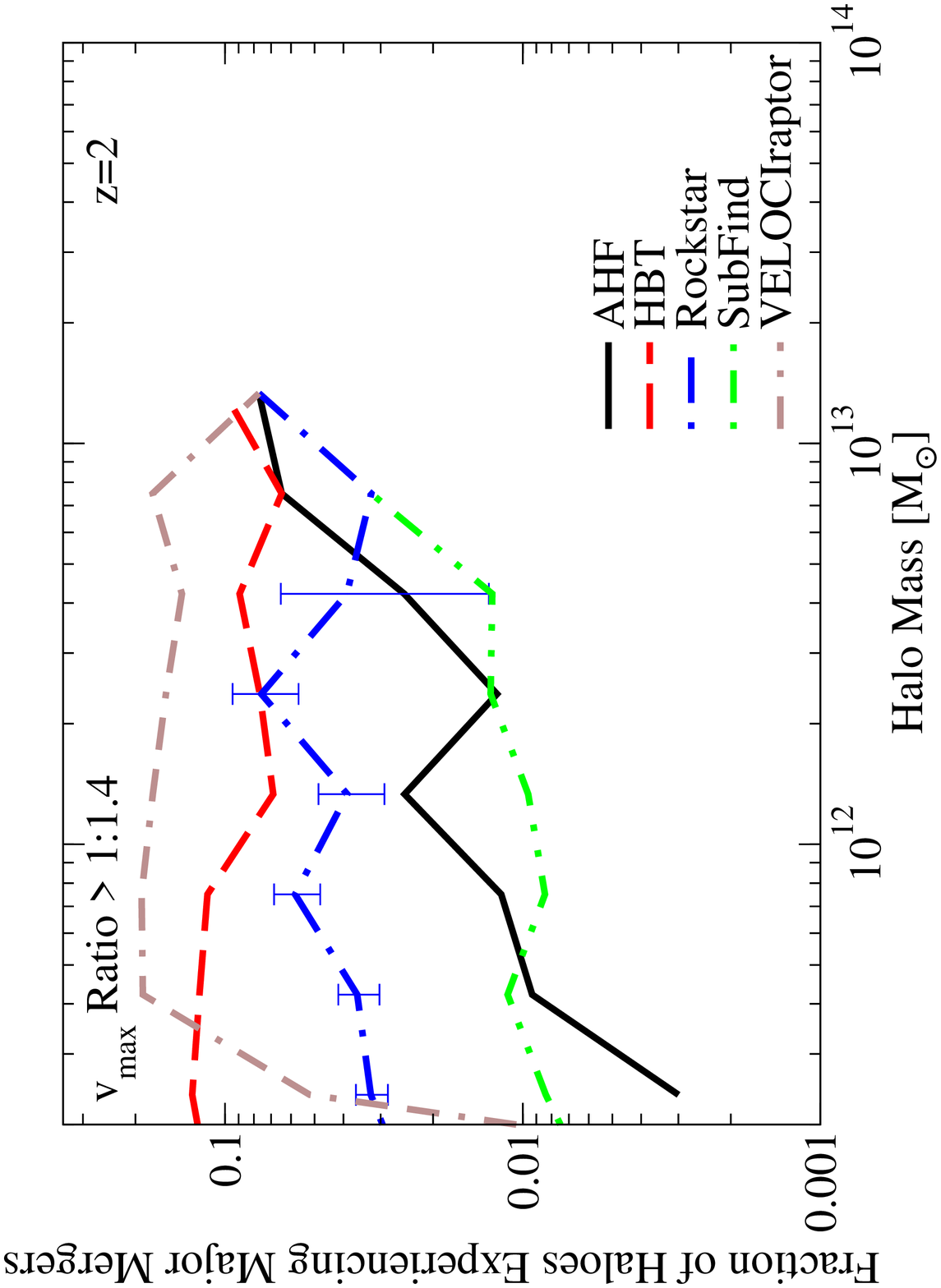}\plotgrace{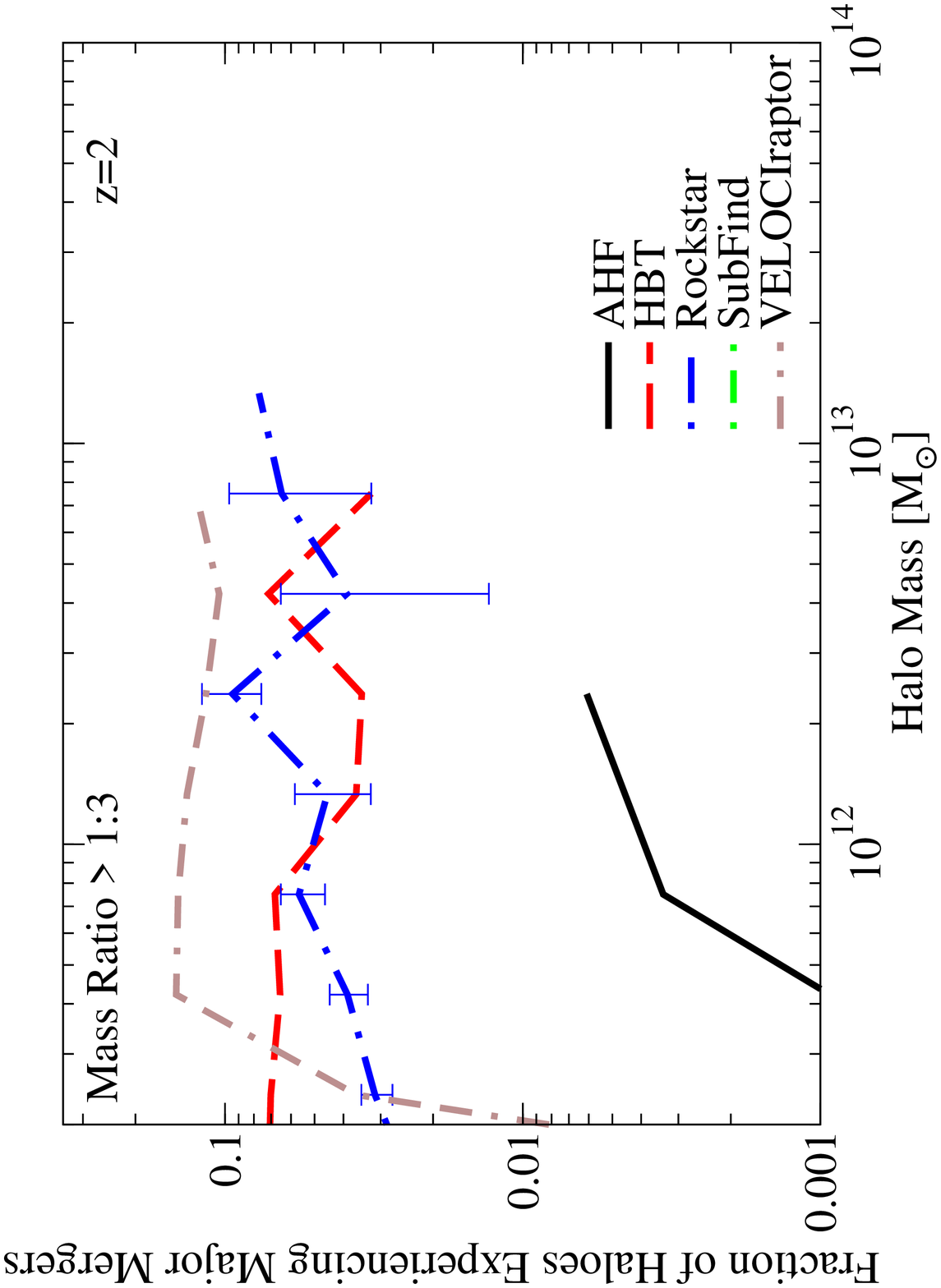}
\caption{Fraction of host haloes undergoing major mergers as a function of mass.  In the \textbf{left} panels, a major merger is defined to be where a subhalo has at least 70\% of the $\vmax$ of its host; this enables fairer comparison with the position-space halo finders.  In the \textbf{right} panels, a major merger is defined to be where a subhalo's mass is at least 33\% that of its host.  Regardless of definition, considerable variation in the reported incidence of major mergers exists at $z=2$.  Some halo finders did not recover major mergers in all mass bins, so their curves above are truncated accordingly.  The error bars show 68\% confidence intervals for \rockstar\ only, to avoid excess clutter; they are not shown in bins where only one major merger was found.   Host halo masses of $1.3\times 10^{11}\Msun$  (left-hand edge of both panels) correspond to 100-particle haloes.  For comparison, the typical collapse mass $M^*$ (i.e., where $\sigma(M^*)=1.686$) is $10^{12.4}\Msun$ at $z=0$ and $10^{9.4}\Msun$ at $z=2$.}
\label{f:frequency}
\end{figure*}

\subsection{Incidence of Major Mergers}

\label{s:cosmo_incidence}

As a gauge of the importance of addressing the issues discussed in this paper, Fig.\ \ref{f:frequency} shows the incidence of major mergers as a function of halo mass and redshift.\footnote{Note that this is separate from the \textit{frequency} of major mergers, discussed in the introduction.  The incidence is the frequency times the average length of time that the merging subhalo remains distinct.}  We define a ``major merger'' in this section to be a subhalo with $\vmax$ at least 70\% of that of its host.  Because $\vmax$ scales as the cube root of mass, this $\vmax$ threshold corresponds approximately to a mass ratio threshold of 1:2.9.  If we had instead defined ``major merger'' in terms of a subhalo mass ratio of 1:3, the position-space halo finders would give very different results from the others (Fig.\ \ref{f:frequency}, right-hand panels) because they often cannot recover full mass profiles for massive subhaloes (\S \ref{s:static}).  Indeed, due to \subfind's conservative approach for assigning particle membership, it finds a factor of over 30 fewer major mergers when using the mass definition as compared to the $\vmax$ definition.  As shown in Fig.\ \ref{f:frequency}, the phase-space and temporal halo finders yield similar incidences for the two definitions.

At $z=0$, with the $\vmax$ definition, there is modest agreement between \ahf, \rockstar, \velociraptor, and \subfind\ that between 2-6\% of haloes are experiencing a major merger; larger haloes are slightly more likely to be undergoing a major merger, as they have later formation times \citep{Wechsler02}.  As noted in \S \ref{s:cosmo_individual}, \hbt\ tracks major mergers for significantly longer than the other finders, with the result that its major merger incidence is elevated with respect to the others.

At $z=2$, there is significant (1 dex) disagreement in the incidence of major mergers, as well as in the change in incidence compared to $z=0$.   For $10^{12}\Msun$ host haloes ($\sim$ 1000 particles), \subfind\ finds lower incidences of major mergers at $z=2$ than at $z=0$ (by 0.4 dex), whereas \ahf\ and \hbt\ find similar incidences at $z=2$ as at $z=0$, and \rockstar\ and \velociraptor\ find higher incidences at $z=2$ than at $z=0$ (by 0.4 dex and 0.8 dex, respectively).  Given the variance even between similar algorithm classes, this likely reflects significant disagreements over where to truncate subhalo mass profiles (as in the static tests in \S \ref{s:static}).  We note that there is apparent convergence in Fig.\ \ref{f:frequency} for the high-mass major merger fraction, but this is illusory: there are not enough haloes in the simulation volume to determine whether or not the halo finders agree above a mass of 5$\times 10^{12}\Msun$.  \ahf\ shows a significant rise in the merger fraction as a function of mass at $z=2$, which may suggest that its algorithm is sensitive to resolution effects.  All other finders follow the theoretical expectation of more self-similar incidences as a function of mass.

We have also briefly investigated the presence of host-subhalo swaps in the merger histories of haloes at $z=0$, using the merger trees generated for \cite{Perez13}.  We find incidences of 0-20\%, which have some dependence on halo mass and halo finder.  However, we find that the merger tree algorithm has a much stronger influence on the number of host-subhalo swaps than any other variable \citep[see also][]{Srisawat13}.  Merger tree algorithms that use all particles contained within haloes (i.e., including substructure)  to match haloes across timesteps are practically immune to host--subhalo swaps (e.g., \textsc{MergerTree} applied to \ahf; \citealt{AHF}), at the expense of discontinuities in halo positions and velocities \citep{Srisawat13}.  The same applies for algorithms which explicitly try to match halo properties across timesteps (e.g., \textsc{Consistent Trees}; \citealt{BehrooziTree}).  Algorithms that use only uniquely-assigned particles (i.e., excluding substructure) are much more vulnerable to host--subhalo swaps, although they do maintain more consistent halo positions and velocities.  Due to the strong dependence on the merger tree algorithm, we postpone further analysis of this to a future paper on the effect of merger tree algorithms on major mergers.

\section{Discussion}
\label{s:discussion}

As we have shown in isolated mergers (\S \ref{s:static} and \S \ref{s:dynamic}), position-space finders cannot accurately recover subhalo masses in major mergers, although recovery of $\vmax$ is not as severely affected.  As shown in \S \ref{s:dynamic}, phase-space finders may perform better when the merger is in its early stages, but also have problems when the merger is nearing completion.  Finally, while temporal algorithms can track halo masses well in the final stages of the merger, the initial masses of the merging haloes may not agree with other approaches (\S \ref{s:cosmo_individual}).  It would therefore seem that some kind of hybrid approach is necessary for best accuracy---e.g., phase space when the merging halo can still reasonably gain mass, and temporal tracking once the merging halo is deep within its host.  Alternately, information from temporal tracking and phase-space could be weighted in a combined metric for particle assignment.

As discussed in \S \ref{s:cosmo_individual}, temporal tracking of particles raises the issue of when two haloes should be considered to have merged.  For smaller mergers, this question is less ambiguous, because tidal stripping removes much of the mass before the merging halo core can sink to the centre.  However, in major mergers, the dynamical friction is such that the merging halo reaches the centre of the host halo before tidal stripping removes all the merging halo's mass.  When the centres of the two haloes meet in position and velocity space, tidal ``stripping'' ceases to be well defined; instead, the profiles of the merging halo and the host halo gradually align until they are indistinguishable.  While some applications may benefit from continuing to track the merging subhalo past this point (e.g., recovery of tidal streams; \citealt{Elahi13}), many others may wish for such subhaloes to be removed or flagged.  A reasonable criterion for removal may be when the phase-space ellipse containing the innermost $N$ particles of the subhalo (where $N\sim 30$) also contains more than $N$ particles from the host or other subhaloes.

Section \S \ref{s:cosmo_incidence} shows that the fraction of haloes undergoing major mergers varies by up to a decade across halo finders at $z=2$, even when the mergers are tagged by $\vmax$ instead of by mass.  As the recovery of the haloes' $\vmax$ prior to the merger is very robust, the main interpretation is that the merger timescales and mass-loss rates differ enormously between different halo finders.  Semi-analytical models which depend on major mergers for aspects of galaxy formation (e.g., morphology/size changes, black hole growth, or star formation triggering) may therefore give very different results when using different halo finders.  Abundance matching and similar empirical models' predictions for the clustering of massive galaxies and the stellar mass content of clusters \citep{Leauthaud12b} may also be affected by the choice of halo finder.  Until there exists a reliable method for determining which subhaloes host visible galaxies \citep[see recent progress in][]{wetzel-09,Watson12,Reddick12,Watson15}, this uncertainty will persist.

\section{Conclusions}
\label{s:conclusions}

We have examined the recovery of host halo and subhalo properties in major mergers across five different halo finders.  Our main findings are summarized as:
\begin{enumerate}
\item Position-space finders recover subhalo positions and velocities well, as long as they can detect the subhalo (\S \ref{s:static}).  The recovered subhalo $\vmax$ is only somewhat biased, but masses are especially difficult to recover accurately (\S \ref{s:static}, \ref{s:dynamic}, \ref{s:cosmological}).
\item Phase-space finders also recover subhalo positions and velocities well, and recover accurate masses in static profile tests (\S \ref{s:static}).  However, in dynamic and cosmological tests, phase differences between the orbiting halo cores and the remaining merger mass can cause large, periodic fluctuations in the recovered subhalo masses (\S \ref{s:dynamic} and \ref{s:cosmological}).  This primarily occurs when the merger is nearing completion.
\item Temporal finders are able to recover positions, velocities, and masses extremely well in isolated merger tests (\S \ref{s:static} and \ref{s:dynamic}).  However, the algorithm tested here (\hbt) does not always reproduce the mass growth history of merging haloes \textit{prior to} the haloes merging (\S \ref{s:cosmo_individual}), when position-space and phase-space finders would be expected to give reliable results.  In addition, while HBT tracks mergers for much longer than other halo finders, the resulting subhaloes are not always distinguishable in phase space from the host particles (\S \ref{s:cosmo_individual}).
\item The fraction of haloes undergoing major mergers is in relative agreement across halo finders at $z=0$ (except for HBT); however, there is strong disagreement ($>1$ dex) in this fraction by $z=2$.
\item All halo finders not using some kind of temporal information show host-subhalo relationship swaps (\S \ref{s:dynamic} and \S \ref{s:cosmo_individual}).  However, the merger tree algorithm employed can to a large degree eliminate this problem (\S \ref{s:cosmo_incidence}).
\end{enumerate}
These findings suggest caution when interpreting results from theoretical models depending on major merger rates at $z > 0$, and they also suggest that future halo finders have ample room for improvement in their treatment of major halo mergers.

\section*{Acknowledgements}

The authors contributed to this paper in the following ways: PB wrote the paper and performed the analysis, with substantial text contributions from AK. AK, FRP \& HL organized the series of workshops where this work was initiated. PB, AK, PJE, JH, YYM, SIM \& CS supplied halo finder data as indicated in \S \ref{s:desc}. SIM generated the mock NFW halo profile, and CS \& DP generated the simulations used. All authors had the opportunity to proofread and comment upon the paper.  This paper uses results from three workshops, including ``Sussing Merger Trees'' (supported from the European Commission’s Framework Programme 7, through the Marie Curie Initial Training Network CosmoComp; PITN-GA-2009-238356), ``Subhaloes Going Notts'' (also supported from EC FP7), and ``Haloes Going MAD'' (supported through the ASTROSIM network of the European Science Foundation; Science Meeting 2910).

Support for PB was provided by a Giacconi Fellowship and an HST Theory grant; program number HSTAR-12159.01-A was provided by NASA through a grant from the Space Telescope Science Institute, which is operated by the Association of Universities for Research in Astronomy, Incorporated, under NASA contract NAS5-26555.  AK is supported by the {\it Ministerio de Econom\'ia y Competitividad} (MINECO) in Spain through grant AYA2012-31101 as well as the Consolider-Ingenio 2010 Programme of the {\it Spanish Ministerio de Ciencia e Innovaci\'on} (MICINN) under grant MultiDark CSD2009-00064. AK also acknowledges support from the {\it Australian Research Council} (ARC) grants DP130100117 and DP140100198. AK further thanks Air for Moon Safari.  HL acknowledges a fellowship from the European Commissions Framework Programme 7, through the Marie Curie Initial Training Network CosmoComp (PITN-GA-2009-238356).  SIM acknowledges the support of the STFC consolidated grant ST/K001000/1 to the astrophysics group at the University of Leicester.  PJE is supported by the SSimPL programme and the Sydney Institute for Astronomy (SIfA), and through the ARC via DP130100117. 

{\footnotesize
\bibliography{master_bib}
}

\label{lastpage}

\end{document}